# *More than just smoke and mirrors*: Gas-phase polaritons for optical control of chemistry

Jane C. Nelson[1], Marissa L. Weichman[1,a]

[1]*Department of Chemistry, Princeton University, Princeton, New Jersey, 08544, USA*

**Abstract**

Gas-phase molecules are a promising platform through which to elucidate the mechanisms of action and scope of polaritons for optical control of chemistry. Polaritons arise from the strong coupling of a dipole-allowed molecular transition with the photonic mode of an optical cavity. There is mounting evidence of modified reactivity under polaritonic conditions; however, the complex condensed-phase environment of most experimental demonstrations impedes mechanistic understanding of this phenomenon. While the gas phase was the playground of early efforts in atomic cavity quantum electrodynamics, we have only recently demonstrated the formation of molecular polaritons under these conditions. Studying the reactivity of isolated gas-phase molecules under strong coupling would eliminate solvent interactions and enable quantum state resolution of reaction progress. In this Perspective, we contextualize recent gas-phase efforts in the field of polariton chemistry and offer a practical guide for experiment design moving forward.

## I. Introduction

Since the invention of the laser, physical chemists have explored the prospect of controlling chemical reactivity with light for practical and intellectual purposes alike.[1–4] Even with decades of progress in laser-driven chemistry, the rapid inter- and intramolecular reorganization of deposited energy in most molecular systems still presents a major obstacle.[5,6] Laser control of chemical behavior has therefore been largely limited to simple gas-phase systems where dephasing and energy redistribution are minimized.[2,7] Recently, evidence of altered solution-phase chemistry under strong light-matter coupling has reinvigorated the dream of optical reaction control under less-than-pristine conditions.[8,9] While new paradigms in

[a)] Author to whom correspondence should be addressed: weichman@princeton.edu

chemical physics have historically been developed in the gas phase before extension to the condensed phase, polariton chemistry has so far skipped this step: the first demonstrations of strong light-matter coupling were performed with atomic gases but gaseous molecular polaritons have not been part of the conversation until our recent work.[10,11] In this Perspective, we review these advances and provide a roadmap for future efforts.

Polaritons are hybrid light-matter states that form when the photonic mode of an optical cavity is brought into resonance with a bright transition of intracavity atoms, molecules, or material.[9,12,13] Under strong coupling conditions, two spectral features appear in the frequency-domain cavity transmission spectrum separated by the Rabi splitting $\Omega_R$ (Fig. 1). When an ensemble of $N$ molecules are coupled to the same cavity mode, the Rabi splitting is observed to scale with $(N/V)^{1/2}$ where $N/V$ is the intracavity molecular number density. These split peaks are dubbed the upper and lower polariton states and can be resolved when the light-matter coupling strength $g = \Omega_R/2$ exceeds both the molecular half-linewidth $\gamma$ and the cavity half-linewidth $\kappa$. In the time domain, strong light-matter coupling indicates that the rate of exchange of photons between the molecules and cavity outcompetes all dissipative processes.

Polariton formation and the emergence of the strong coupling regime are often introduced using the language of cavity quantum electrodynamics (cQED), in which the discrete states of a quantum emitter hybridize with a quantized mode of light.[14] In the simplest case, the Jaynes-Cummings (JC) model treats the interaction of a single two-level emitter with a harmonic photonic mode, considering only the first excitation manifold of the coupled light-matter system.[15,16] When the JC Hamiltonian is diagonalized, the upper and lower polaritons emerge as eigenstates corresponding to symmetric and antisymmetric linear combinations of photonic and material excitations. The Tavis-Cummings (TC) model extends the JC model to the collective strong coupling regime by accounting for the mixing of one cavity mode with the totally-symmetric collective bright state of $N$ intracavity two-level emitters.[17] Upper and lower polariton states again emerge, separated in frequency by a collective Rabi splitting that scales with $N^{1/2}$. In addition to the two polaritonic eigenstates, $N - 1$ so-called "dark states" are left at the original energy of the uncoupled emitters. The dark states are non-totally symmetric linear combinations of emitter states that carry no photonic character and are therefore optically forbidden (so long as one neglects heterogeneity in the ensemble of emitters).[18]

cQED models provide a convenient conceptual description of polaritons as hybrid light-matter quasiparticles and correctly predict the Rabi splitting and its collective scaling for intracavity ensembles. However, experimental implementations of the strong light-matter coupling regime feature complexity not captured by the JC or TC models: molecules are not two-level systems and they exhibit finite excited-state lifetimes as well as energetic and orientational disorder, while optical cavities have finite lifetimes due to photonic losses and feature a geometry-dependent spectrum of longitudinal and transverse modes. In addition, most experiments use classical light sources to probe cavity transmission spectra, operating nowhere near the few-photon regime of quantum optics. As a result, many experimental groups treat cavity spectra using classical simulations which can more easily account for the aforementioned practical factors.

Classically, the strong coupling regime arises from the dispersion of light in a cavity filled with dielectric material. In particular, we consider the self-interference of an incident electromagnetic field as it propagates through a Fabry-Pérot (FP) cavity composed of two identical mirrors spaced by length $L$ and containing an intracavity medium with a frequency-dependent refractive index $n(\nu)$ and absorption coefficient $\alpha(\nu)$. The fractional intensity of light transmitted through such a structure is given by:[9,19,20]

$$\frac{I_T(\nu)}{I_0} = \frac{T^2 e^{-\alpha(\nu)L}}{1+R^2 e^{-2\alpha(\nu)L} - 2Re^{-\alpha(\nu)L}\cos\left(\frac{4\pi L n(\nu)\nu}{c}\right)} \quad (1)$$

where $R$ and $T$ are the reflection and transmission probabilities for each mirror with $R + T = 1$ for lossless mirrors. The widely-used transfer matrix method uses the same principles of classical wave interference as Eq. (1) to treat cavity structures containing layered media.[9,13,21] Peaks in the cavity transmission spectrum appear through constructive interference of light traveling in the cavity at frequencies where the round-trip phase shift ($\delta\phi = 4\pi L n(\nu)\nu/c$) is equal to an integer multiple of $2\pi$. When $n(\nu)$ is constant, cavity transmission peaks are evenly spaced by the cavity free spectral range (FSR). Near a strongly-absorbing molecular transition, on the other hand, $n(\nu)$ takes on a dispersive line shape and gives rise to additional transmission fringes whose frequencies coincide with the polaritonic states predicted by cQED. Recent theoretical work has drawn an explicit connection between the cQED and classical optics results in the limit that the number of intracavity absorbers $N \to \infty$, explaining why both descriptions are successful in capturing experimental data.[22,23]

While the community has focused on strong light-matter coupling in condensed-phase systems for the past 15 years, the original demonstrations of cQED were carried out in atomic gases. In the early 1980s, Haroche and coworkers observed the resonant enhancement of spontaneous emission under *weak* coupling of sodium Rydberg atoms to a microwave Fabry-Pérot resonator,[24] a manifestation of the long-predicted Purcell effect.[25,26] Shortly thereafter, Haroche's team reported *strong* cavity coupling in the same $N$-atom system,[27–29] measuring the excited state Rydberg population in the time domain via field ionization and observing collective Rabi oscillations with frequency scaling as $N^{1/2}$. Further, Haroche showed reversible spontaneous emission under strong coupling as the cavity recycles photons back to the atomic ensemble faster than excitations dissipate.[28] Within the same decade, Kimble and coworkers harnessed improvements in mirror coatings to strongly couple electronic transitions of sodium atoms at optical wavelengths.[30] Kimble's group then advanced towards the single-atom limit, achieving the first experimental demonstration of the JC model with a cesium atom.[31–33] Kimble and coworkers also introduced metrics for strongly-coupled systems via frequency-domain cavity transmission spectroscopy, reporting both the splitting of the coupled system's normal modes and dispersion of these modes upon detuning the cavity from resonance.[30] To this day, mode splitting and dispersion in cavity transmission or reflection spectra remain the prime experimental hallmarks of strong coupling.

Polariton formation has subsequently been explored in a variety of media. Inspired by the early atomic demonstrations, solid-state physicists reported strong light-matter coupling in semiconductor quantum wells in 1992[34] and launched the field of inorganic exciton-polaritons.[35] In 1998, Lidzey and coworkers made the leap to create exciton-polaritons in organic molecular materials.[36] Solid-state exciton-polaritons have since remained an active research topic with Bose-Einstein condensation and polariton lasing being of particular interest.[37–39] In the 2010s, Ebbesen and coworkers led a new charge to examine the chemical behavior of molecular polaritons,[40] focusing initially on electronic strong coupling (ESC) in molecular dyes and the resulting impact on photochemistry and fluorescence.[41–43] In 2015, Ebbesen[44] and Simpkins[45] reported parallel implementations of vibrational strong coupling (VSC) by coupling infrared-active carbonyl modes of polymer films to micron-scale planar FP cavities. Both teams subsequently demonstrated that solution-phase molecules with strong and narrow absorption bands could be cavity-coupled in synthesis-compatible microfluidic devices.[46,47] Ebbesen and coworkers took the next step to examine chemical reactivity under VSC, reporting *slowed* kinetics of a thermal liquid-phase silane deprotection reaction in a microcavity in 2016.[48] In the near-decade since, several groups have worked to understand the scope and reproducibility of both ground-state chemistry under VSC[8,9,13,49–55] and photochemistry under ESC.[56–60] Despite a vibrant research community and a growing body of experimental work in polariton chemistry, the field is still in search of mechanistic understanding and predictive capabilities. Solvation effects and rapid energy dissipation complicate data analysis and interpretation in liquid-phase chemical kinetics and dynamics.[6,61] Meanwhile, theoretical efforts face the challenge of simultaneously treating the experimentally-relevant collective coupling of $N > 10^6$ molecules, disorder, complex reactive surfaces, and realistic lossy multimode cavities.[14,62,63]

Here, we do not aim to provide a comprehensive review of all prior work and proposed mechanisms for cavity-altered chemistry, and refer interested readers to several recent reviews of this emerging and lively field.[9,12–14,62,63] Instead, we lay out a case for how gas-phase molecular polariton experiments may clarify the state and scope of polariton chemistry. Gas-phase molecules feature resolved quantum states and well-characterized potential energy surfaces that may prove useful in identifying signatures and validating emerging hypotheses for cavity-mediated dynamics. For example, a leading hypothesis for chemistry under VSC is that the cavity may mediate intramolecular vibrational energy redistribution (IVR) among the ensemble of strongly-coupled intracavity molecules.[14,54,63–69] In free space, IVR can cause a selectively-excited vibration to decay into the manifold of nearby states on ultrafast timescales.[70] If operative, a cavity-IVR mechanism could drain energy out of a strongly-coupled unimolecular or bimolecular reaction coordinate into orthogonal molecular degrees of freedom or into the lossy photonic mode; this would serve to slow reaction rates under VSC, as has indeed been observed in seminal experiments.[48,53] At present, just one report has provided evidence of enhanced IVR in a unimolecular isomerization under VSC.[64] Additional experiments are necessary to validate the cavity-IVR mechanism and other emerging hypotheses in a wide range of chemical systems, including bimolecular reactions. We argue here that the gas phase is a sensible domain in which to conduct this work, as the historical proving ground for chemical physics which the polariton community has not returned to since the original atomic cQED experiments.

In this Perspective, we report on the status and future prospects of gas-phase molecular polaritons. We lay out important experimental considerations for gas-phase polaritons and review our group's recent demonstrations in Sec. II. In Sec. III, we describe our procedure to evaluate new molecular candidates using the classical optics framework described above. We propose several next-generation species for both vibrational and electronic gas-phase strong coupling in Sec. IV, highlighting the most compelling prospects for studies of reactivity and photophysics. Finally, in Sec. V, we consider future challenges and envision how gas-phase strong coupling efforts may prove a crucial step in bridging the experiment-theory gap in cavity chemistry.

## II. Laying the Groundwork for Gas-Phase Molecular Polaritons

Here, we briefly review the infrastructure our group has introduced to cavity couple gas-phase molecules.[10,11] Reaching the strong coupling regime in diffuse gases requires some thought given the $(N/V)^{1/2}$ dependence of the collective Rabi splitting; simply increasing the gas pressure can lead to counter-productive pressure broadening of molecular transitions.[13] Instead, we operate in the Doppler-broadened regime at relatively low temperatures and pressures, aiming to simultaneously minimize molecular linewidths and maximize $\Omega_R$. We use collisional cooling in a home-built cryogenic buffer gas cell (CBGC) to prepare a cold, dense intracavity sample (Fig.

2a). Through cooling, we narrow Doppler line shapes and reduce the rovibrational partition function to condense population into low-lying states. To perform cavity-coupling *in situ*, we surround the CBGC with a plano-concave two-mirror FP optical cavity whose transmission we probe with a tunable narrow-band mid-infrared continuous-wave (cw) laser.

Our open centimeter-scale cavity design enables careful tunability of parameters and distinguishes our gas-phase apparatus from the micron-scale microfluidic cavities more commonly used for VSC.[9] For a given experiment, we choose a cavity geometry to match the photonic mode linewidth to that of the targeted molecular transition and to ensure the cavity mode spacing is large enough for clean, state-specific molecular coupling conditions. In practice, we have direct control over:

- the cavity length, $L$, which sets the free spectral range as FSR = $c/2nL$, where $c$ is the speed of light and $n$ is the intracavity refractive index;
- the reflectivity of each mirror, $R$, which sets the cavity finesse according to $\mathcal{F} = \pi R^{1/2}/(1-R)$, assuming the two mirrors are identical; practically, this parameter is used in combination with the cavity length to determine the photonic mode linewidth as $\Delta\nu = \text{FSR}/\mathcal{F}$;
- and, the mirror radius of curvature (ROC) which determines the spectrum of transverse Gaussian spatial modes supported by a plano-concave FP cavity.

A representative experimental cavity transmission spectrum is plotted in red in Fig. 2b for a near-confocal cavity with $L$ = 8.36 cm, ROC = −8.36 cm, and $\mathcal{F}$ = 24. We mount one cavity mirror on a piezo-electric (pzt) chip to allow remote detuning of the cavity length. We stabilize the cavity length with active feedback on the pzt to ensure that cavity-coupling conditions remain consistent throughout an experiment.

We use rovibrational transitions of methane ($CH_4$) for an initial demonstration of gas-phase strong coupling.[10] $CH_4$ is a convenient target with high symmetry, bright line strengths, and precedence in the mode-specific chemistry literature.[4,71] We target an individual rovibrational transition in the R-branch of the $v_3$ asymmetric C−H stretching band (blue trace in Fig. 2b). At low temperature and pressure, the three $A_2(0)$, $F_2(0)$, and $F_1(0)$ symmetry components of the $v_3$, $J = 3 \rightarrow 4$ rovibrational transition can be resolved at 3057.687423, 3057.726496, and 3057.760735 $cm^{-1}$, respectively.[72] We record the full-width at half-maximum (fwhm) linewidth of these peaks as 180 MHz (0.006 $cm^{-1}$) in the CBGC, consistent with Doppler broadening at 120 K. To perform strong coupling experiments, we stabilize the cavity length so one photonic mode is resonant with a targeted molecular transition, then flow $CH_4$ into the cell while monitoring the cavity transmission spectrum. Split polaritonic features appear at sufficiently high molecular flow rates; a representative strong coupling experiment targeting resonance with the $A_2(0)$ band is given in the purple trace in Fig. 2b. All experimental traces are well reproduced by the classical cavity transmission expression in Eq. (1).

Following the initial demonstration of this gas-phase strong coupling platform, we explore the tunability of both cavity and molecular experimental parameters.[11] We continue to work with the rovibrational $v_3$, $J = 3 \rightarrow 4$ $A_2(0)$ band of $CH_4$, first testing a wider range of molecular conditions. At molecular number densities where the collective Rabi splitting approaches the cavity FSR, we observe that nested polariton peaks emerge from the off-resonant coupling of the targeted molecular transition with adjacent cavity modes (Fig. 2c). The additional peaks in the cavity transmission spectrum become even more numerous at higher flow rates where the collective Rabi splitting approaches the spacing between molecular states. These emergent features represent admixtures of multiple molecular transitions and multiple photonic modes; their light-matter composition could be quantified with Hopfield coefficient analysis.[9,63] We also show that strong coupling of $CH_4$ is possible at room temperature, though the increased room temperature rovibrational partition function and Doppler broadening lower the absorption cross section of the targeted transition. We therefore need an order of magnitude higher molecular number density to reach the same room temperature Rabi splitting as in a 120 K sample, and only a small fraction of these warm molecules are in the correct lower state for resonant cavity-coupling.[11] Regardless, working with room temperature samples greatly relaxes experimental demands and lowers the barrier for others in the community to create and study gas-phase molecular polaritons.

We also systematically tune the cavity parameters and examine the effects on polariton formation.[11] Evaluating cavity transmission spectra as a function of cavity finesse, we find that polariton linewidths track with the changing cavity mode linewidth while the Rabi splitting is unchanged, consistent with the known behavior of inhomogeneously-disordered samples under strong coupling.[13,73] Tuning the FSR via the cavity length enables further control of the photonic modes participating in strong coupling. By choosing an FSR commensurate with the spacing between rovibrational molecular peaks, we can reach a multi-state coupling condition where two transitions are simultaneously cavity-coupled (Fig. 2d).

Extending this apparatus to strongly couple molecules other than $CH_4$ is straightforward. All that is required is (a) a tunable cw laser to measure cavity transmission in the targeted spectral region; (b) cavity mirrors with reflectivity in the targeted spectral region, though the gold-coated mirrors we have used so far are quite general purpose; and (c) a cavity geometry with an FSR and linewidth designed to couple cleanly to the desired molecular transition. We discuss some practicalities in screening next-generation candidates in Sec. III, below.

**III. Evaluating Next-Generation Candidates for Gas-Phase Strong Coupling**

Here we lay out our considerations for extending our platform to molecules of interest in polariton reactivity studies. We aim to provide a roadmap for designing gas-phase polariton experiments that target molecules with well-studied reactivity in free space, particularly those with unimolecular or IR-driven chemistry. We must carefully consider the experimental feasibility of target systems, a step often overlooked in theoretical proposals in the literature. Specifically:

1. Prospective molecules must first and foremost be amenable to strong coupling under physically realizable conditions. The collective Rabi splitting – which has to exceed the molecular and cavity linewidths to reach the strong coupling regime – is proportional to the molecular transition dipole and the square-root of the number density. A target molecule must have both sufficiently bright optical transitions and a large vapor pressure so reasonable gas-phase number densities are experimentally accessible.

2. The molecule's fully-resolved rovibrational or vibronic structure must also be considered. In the condensed phase, fine structure is washed out and an entire vibrational or electronic band can be coupled at once. In the gas phase, we can couple individual quantum states with proper cavity design provided that we account for the temperature- and pressure-dependent absorption line shape of the targeted transition.

3. Finally, we must choose systems with direct readouts of reaction dynamics and which are likely to feature cavity-modulated behavior. We also have to consider whether it is possible to study a reaction at the high molecular number densities required for strong coupling and evaluate possible reactant or product detection schemes.

We will discuss this last point further for various systems in Sec. IV below. Addressing the first two points is more straightforward: we can evaluate prospects for strong coupling specific molecular transitions by simulating the transmission of light through a cavity containing the target species with the classical optics treatment described in Sec. I.[10,11,19,20] The most critical components of these simulations are the molecular absorption coefficient, $\alpha(\nu)$ [cm$^{-1}$] and the real part of the refractive index, $n(\nu)$. $\alpha(\nu)$ is typically determined from the absorption cross section, $\sigma(\nu)$ [cm$^2$/molecule], according to $\alpha(\nu) = \sigma(\nu) \cdot N/V$ and must be evaluated in the spectral range of interest under experimentally relevant number density, Doppler broadening, and pressure broadening conditions. $n(\nu)$ can then be derived from $\alpha(\nu)$ using the Kramers-Kronig relation. Finally, we input $\alpha(\nu)$ and $n(\nu)$ into Eq. (1) along with experimentally reasonable parameters for mirror reflectivity and transmission $(R,T)$ and cavity length $(L)$ to simulate transmission spectra. For most simulations provided below in Sec. IV, we assume a cavity geometry similar to that used in our previous work.[10,11] Unless otherwise noted, we use a representative non-confocal $L = 5$ cm optical cavity with $R = 90\%$ mirrors. This geometry features a mode spacing of FSR = 3000 MHz (0.1 cm$^{-1}$), a mode linewidth of $\Delta\nu = 100$ MHz (0.003 cm$^{-1}$) fwhm, and a finesse of $\mathcal{F} = 30$. We confirm strong coupling prospects for each system by targeting resonance between a cavity mode and a selected molecular transition for various number densities and determine the conditions for which polaritonic splittings first appear in the transmission spectra.

It is straightforward to perform these simulations when high-resolution reference data exist for the absorption cross section $\sigma(\nu)$. In the best-case scenario, a line list is available from the HITRAN database[72] which can be opened directly in the PGOPHER software package.[74] The process is more involved when line lists or high-resolution experimental reference data are not available, the target species has many isomers, or the system features intrinsic broadening. In these cases, the temperature and pressure dependence of $\sigma(\nu)$ may be less certain. When lacking a line list, we build a rovibrational or vibronic model of the system in PGOPHER using spectroscopic constants derived from Gaussian 16 calculations[75] and any available experimental spectroscopic constants. The PGOPHER documentation[74] and tutorials from Sprague[76] and Wilhelm *et al.*[77] are helpful in this process. We then use the PGOPHER spectrum fitting tools to fit the available experimental reference spectra. Once we have constructed a reasonable simulation, we set the temperature, Doppler broadening, and pressure broadening in PGOPHER to calculate $\sigma(\nu)$ under conditions relevant to our CBGC experiments and convert to the absorption coefficient $\alpha(\nu)$ for various molecular number densities.

Table I provides a summary of the gas-phase molecular transitions that we examine for strong coupling prospects in this work, which we discuss in Sec. IV below. See Sec. SI of the Supplementary Material (SM) for more details and Table SI for specific simulation parameters and sources of reference data.

### IV. Prospective Molecular Gases for Polariton Chemistry

We now report specific prospects for strong cavity coupling of quantum-state-resolved transitions in molecules of interest. We target rovibrational strong coupling (RVSC) in molecules that undergo gas-phase bimolecular H-abstraction reactions and unimolecular isomerization and photodissociation. We also explore possibilities for gas-phase electronic strong coupling at UV/visible wavelengths. In our evaluation of RVSC candidates, we place a particular emphasis on systems whose behavior may inform the cavity-mediated IVR hypothesis for polariton chemistry introduced in Sec. I. We therefore target (a) reactions with established IR-driven reactivity where we can examine if intracavity rates and product branching are modifiable via cavity-IVR; *and* (b) molecules that natively feature IVR where we might resolve vibrational energy distributions and screen for cavity-induced changes.

*A. Bimolecular reactions of small hydrocarbons*

The bimolecular hydrogen (H) abstraction reactions of small organic molecules with reactive radicals are benchmark systems for laser-driven chemistry[2,71] and feature simple, well-characterized reactive surfaces for theoretical analysis. The reactions of methane ($CH_4$) and acetylene ($C_2H_2$) with the hydroxyl radical (OH)[78,79] and other small radicals have been the subject of decades of research.[80] We have already demonstrated RVSC of $CH_4$;[10,11] here we show that acetylene ($C_2H_2$) has similar prospects. We consider the rovibrational transitions of room-temperature $C_2H_2$ in the $\nu_3$ C−H stretching band near 3 μm using line lists available in HITRAN. We predict the onset of strong coupling for $N/V = 3\times10^{15}$ cm$^{-3}$ (Fig. 3, Table I). As $C_2H_2$ is a commercially available gas, reaching this number density is not a practical concern. Comparable Rabi splittings are accessible for the same number densities of $CH_4$ and $C_2H_2$, though $C_2H_2$ features stronger polaritonic transmission features due to its narrower Doppler linewidth and therefore weaker absorption at polariton frequencies (Fig. 3). We expect cooling of $C_2H_2$ to reduce the number density necessary to reach strong coupling, as we saw for $CH_4$.[11]

The H-abstraction reactions of strongly-coupled $CH_4$ or $C_2H_2$ with OH can be monitored with pulsed laser photolysis-laser-induced fluorescence (PLP-LIF). In this approach, one would prepare OH reactants via PLP of hydrogen peroxide precursor and track the subsequent reactive loss of OH using LIF.[78,79,81] Derived reaction rates can be compared between on- and off-resonance intracavity experiments and extracavity controls. The mechanisms of cavity-altered bimolecular reactions remain poorly understood but the relevant experimental literature, if anything, suggests that they will feature slower kinetics under VSC of reactants.[48,53] We therefore hypothesize that one might observe slowed reaction rates for intracavity gas-phase H-abstraction. In any event, these systems will be a natural platform to test theories like the cavity-IVR hypothesis introduced in Sec. I, which posits that cavity-coupling can impede reaction progress by introducing a new channel to drain vibrational energy out of the reaction coordinate. Adding a detection channel to probe the distribution of post-reaction product vibrational states could explicitly illuminate intracavity vibrational energy redistribution and more directly test the cavity-IVR hypothesis. Preparation of vibrationally-excited reagents is another natural direction; IR-pumping of both reactants in the $CH_4$ + OH system has been examined in the extracavity literature.[82–84] Performing similar experiments under RVSC of $CH_4$ would provide a direct probe of cavity-mediated relaxation pathways that may funnel vibrational energy out of the pumped reaction coordinate.

The bimolecular reactions of larger organic molecules that sustain significant inter-mode vibrational couplings are also of interest for gas-phase polariton chemistry studies and could enable further insight into the cavity-IVR hypothesis. Methanol ($CH_3OH$) has similar reactivity to methane[85,86] but features more extensive torsion-vibration couplings and rapid IVR.[87–89] Similarly, extending $C_2H_2$ to longer acetylenic chains (e.g. 1-propyne, 1-butyne, 1-pentyne) yields denser manifolds of vibrational states and increased coupling between modes.[90,91] Unfortunately, we predict that it will be practically challenging to reach the strong coupling regime in these species, as the same intramolecular couplings of interest cause spectral broadening and congestion that prohibit clean, state-specific resolution of polaritonic features. For example, while transitions in the C−O stretching band of methanol near 10 μm are accessible to cavity-coupling, the intrinsic congestion of the C−H stretching region near 3 μm[92] will prohibit rovibrationally-resolved cavity-coupling (Fig. S1). We note that state-specific resolution of coupling conditions is one of the unique strengths of working with gas-phase molecules; it may make more sense to rely on existing condensed-phase microcavity architectures when targeting broad or unresolved bands. Balancing these competing considerations is a challenge for future work, as we discuss further in Sec. IV D.

*B. Unimolecular isomerization: nitrous acid and butadiene*

Unimolecular isomerization reactions are compelling systems for polariton chemistry as they typically depend on IVR to funnel vibrational excitation into the isomerization coordinate. The *cis-trans* isomerization of nitrous acid (HONO) has been targeted by several theoretical VSC studies.[68,93,94] IR-driven *cis-trans* HONO isomerization was first demonstrated in 1963[95] and has been studied since.[96–101] This process can be driven by pumping the ~3400 cm$^{-1}$ $\nu_1$ O−H stretch of *cis*-HONO despite the reaction coordinate being more similar to the low-frequency $\nu_6$ torsional mode.[97] Vibrational excitation of the O−H stretching coordinate must therefore redistribute into torsional motion for the reaction to progress. This IR-driven isomerization has only been observed in inert gas matrices where coupling to the matrix likely drives the redistribution of vibrational energy into the isomerization coordinate.[96] Still, it is worth considering whether cavity coupling could mediate IVR in gas-phase HONO, especially given the theoretical interest in this system.

We first evaluate strong coupling prospects for the brightest vibrational mode of the dominant *trans*-HONO isomer: the $\nu_3$ HON bending mode (Fig. 4). We build a PGOPHER model for HONO using spectroscopic constants from Gaussian calculations and the literature[97,99,102–105] in order to fit the available room-temperature data[72,106] (red trace in Fig. 4b). The features in the absorption cross section grow significantly narrower and brighter at 100 K (blue trace in Fig. 4b). We identify the $\nu_3$, $J = 9\rightarrow10$ transition at 1271.218 cm$^{-1}$ as an optimal rovibrational transition for cavity-coupling. We find that strong coupling of this transition is possible only with $N/V \geq 3\times10^{15}$ cm$^{-3}$ (Fig. 4c, Table 1). Unfortunately, this number density is unlikely to be realized in the gas phase; the highest HONO number densities reported[99] do not exceed $8.4\times10^{13}$ cm$^{-3}$ and contamination with NO$_x$ byproducts is unavoidable even in dilute HONO samples.[99,107–111] This source issue is likely prohibitive for achieving RVSC in the $\nu_3$ band of the *trans*-HONO or any weaker vibrational transitions of either isomer. To simulate the *trans*-HONO absorption cross section accurately one must account for the changing isomer equilibrium between room temperature and 100 K. Because *trans*-HONO is the thermodynamic ground state we do expect a larger contribution from the *trans*-HONO population at low temperatures. We examine this further in Sec. SIII of the SM but find that this consideration does not relax the inaccessibly high *trans*-HONO number density required for strong coupling.

Though RVSC of HONO is likely not realistic, there are other promising candidates for gas-phase isomerization. Cavity-altered isomerization of 1,3-butadiene (C$_4$H$_6$) has been the subject of recent theoretical work.[112] The planar *trans*-C$_4$H$_6$ isomer composes 95% of gaseous samples at room temperature; rotation about the central C−C bond leads to a non-planar *gauche* structure (Fig. 5a).[113] The ultraviolet photochemistry of C$_4$H$_6$ has been studied for decades[114–117] and IR-driven photoisomerization should be possible given the ~2000 cm$^{-1}$ torsional barrier in the electronic ground state.[118] We evaluate prospects for cavity-coupling the $\nu_{11}$ CH$_2$ wagging mode of *trans*-C$_4$H$_6$ which is the most strongly-absorbing band at room temperature (Fig. 5, Table I). We build a C$_4$H$_6$ model in PGOPHER based on spectroscopic constants from calculations and literature[119–124] and fit it to the available experimental data[72,106] (red trace in Fig. 5b). We neglect the temperature-dependence of the *trans-gauche* isomer equilibrium which would lead to at most a 5% increase in the absorption cross section of *trans*-C$_4$H$_6$ at low temperatures. We inspect the 100 K simulated spectrum (blue trace in Fig. 5b) and select overlapping *trans*-$C_4H_6$ $\nu_{11}$, $J = 8\to7$ and $J = 30\to29$ R-branch transitions at 894.2307 cm$^{-1}$ and 894.2316 cm$^{-1}$ for cavity coupling. We find that the strong coupling regime is accessible for $N/V \geq 1\times10^{16}$ cm$^{-3}$ (Fig. 5c). $C_4H_6$ is a commercially available gas, so achieving these number densities is straightforward. Investigating the IR-driven chemistry of $C_4H_6$ under strong coupling could provide a platform to explore the cavity-IVR hypothesis. It may also be of interest to study IVR-mediated excited-state photochemistry of $C_4H_6$ under RVSC, including photoisomerization and photolysis.

*C. Photodissociation of ozone under RVSC*

We now explore cavity-coupling prospects for ozone ($O_3$) which features rich and well-studied photochemistry[125–127] and is relevant to recent theoretical predictions for cavity-IVR mediated photodissociation of triatomic molecules.[65,66] Here, we consider RVSC in ground-state $O_3$; we discuss possibilities for ESC of this system in Sec. IV D below. Making use of the HITRAN line list for the infrared spectrum of $O_3$, we target the brightest $\nu_3$ asymmetric stretching mode and select the $\nu_3$, $J = 7\to8$ transition at 1048.0695 cm$^{-1}$ for coupling (Fig. 6, Table I). We predict RVSC of $O_3$ to be possible with $N/V = 3\times10^{15}$ cm$^{-3}$ at room temperature and $N/V = 9\times10^{14}$ cm$^{-3}$ at 100 K (Fig. 6bc); these number densities are accessible with commercial ozone generators.

$O_3$ photodissociation under RVSC may be a useful system in which to test the cavity-IVR hypothesis. The theoretical work of Wang *et al.*[65,66] predicts that generalized bent triatomic molecules resist photodissociation under strong (nearly ultrastrong) coupling to the bending vibrational mode. However, there are some important differences between the model system considered by Wang *et al.* and our proposed gas-phase strong coupling of $O_3$. Here, we consider strong coupling of a single rovibrational transition of the asymmetric stretching band of $O_3$. Wang *et al.*, on the other hand, do not account for rotational structure and assume the entire vibrational oscillator strength is concentrated in a single transition. In addition, Wang *et al.* consider coupling to the bending mode of their triatomic model system while we find that strong coupling of the bending mode of $O_3$ is not practical given this band's tiny absorption cross section (Fig. 6a). Studying photodissociation of $O_3$ under RVSC of the asymmetric stretch may still provide a useful step towards experimental tests of theory. It may also be worth exploring RVSC of heteronuclear triatomics (e.g. HOCl, OCS, or HCN) featuring brighter bending modes. Studying dissociation of any of these species following electronic or multiphoton vibrational photoexcitation under strong coupling could aid in understanding cavity mediation of vibrational energy redistribution.

*D. Electronic strong coupling of molecular gases: iodine and ozone*

Electronic strong coupling of molecular gases has not yet been demonstrated but should be feasible. Here, we consider species with narrow, well-resolved vibronic transitions so molecular linewidths are similar to centimeter-scale FP cavity linewidths. Molecular iodine ($I_2$) is one promising candidate. We build a model of the B−X band of $I_2$ near 538 nm (18600 cm$^{-1}$) in PGOPHER using literature constants[128–133] to fit room temperature experimental absorption cross section data (Fig. 7a).[134] Our model accounts for transitions from the $v_{initial}$ = 0, 1, 2, and 3 vibrational levels of the ground electronic state to the $v_{final}$ = 0 to 70 vibrational levels of the excited B state in order to recover the experimental Franck-Condon envelope in the region of maximum absorption. We additionally account for the underlying $I_2$ hyperfine structure[132,135] which manifests in asymmetric line shapes under low pressure, Doppler-broadened conditions.

We confirm that this model reproduces the known state-resolved rovibronic structure of $I_2$ (Fig. 7b).[128] Finally, we simulate room temperature cavity-coupling of the B−X, $v_1 = 0 \rightarrow 29$, $J = 45 \rightarrow 46$ rovibronic transition lying near 18599.1 cm$^{-1}$ (Fig. 7c, Table I). We predict the onset of ESC for $N/V = 5 \times 10^{15}$ cm$^{-3}$. Solid iodine has a room-temperature vapor pressure[134] of 0.23 torr which corresponds to $N/V \sim 7.4 \times 10^{15}$ cm$^{-3}$ so the strong coupling regime should be accessible with the vapors of room-temperature or gently-heated solid $I_2$. Fluorescence and photodissociation yields of $I_2$ could be examined as a function of cavity-coupling conditions.

We finally consider if ESC of a broad, unresolved electronic band is possible in the gas phase. Coupling a broad molecular transition to a single photonic mode demands a cavity with a large FSR and therefore a cavity length more typical of condensed-phase nanocavities ($L < 1$ μm). To give a concrete example, we consider coupling the B−X electronic band of ozone near 254 nm (Fig. S2, Table 1).[136] Pumping this band leads to $O_3 \rightarrow O + O_2$ photodissociation with interesting wavelength-dependence of photoproduct electronic states;[125–127] one could study these channels under ESC. Unfortunately, the $O_3$ B−X band is simply not bright enough for strong coupling given its significant linewidth; we find that strong coupling only emerges for $N/V \geq 3 \times 10^{21}$ cm$^{-3}$ even in a wavelength-scale cavity with $L = 254$ nm (Fig. S2b). For context, this number density corresponds to a physically unrealizable pressure of 122 atm at room temperature, assuming the sample behaves as an ideal gas.

We therefore conclude that molecules with well-resolved vibronic bands and correspondingly narrower natural linewidths are better candidates for gas-phase ESC. Quantum-state resolution is a major strength of gas-phase polaritonics and careful system selection will be central to future work.

## V. Future Challenges and Conclusions

As polariton chemistry continues into its second decade, gas-phase experiments may provide a bridge between condensed-phase demonstrations and theoretical understanding. We have argued here that many exciting proposals for cavity-altered reaction dynamics are well-suited for study in isolated gas-phase molecules. There are practical experimental benefits: generating polaritons in open plano-concave Fabry-Pérot cavities enables unique flexibility and control of both molecular and photonic degrees of freedom. We conclude this Perspective by highlighting some of these opportunities, as well as future challenges.

*Orthogonal optical access.* The relatively long centimeter-scale pathlengths of gas-phase FP cavities provide the opportunity for optical access orthogonal to the cavity axis, which is considerably harder to implement in microcavities. This access will allow for initiation and interrogation of intracavity processes without confounding optical filtering effects. For instance, the hypothetical intracavity PLP-LIF experiments discussed in Sec. IV A could be implemented with PLP (pump) and LIF (probe) beams aligned along the cavity axis and fluorescence detected orthogonally. One could also consider orthogonally-aligned pump (or probe) beams, though one must take case to ensure that the molecules addressed along the orthogonal axis belong to the cavity-coupled ensemble. Designing the cavity length and beam diameter to be comparable in size (e.g. a few mm) would enable the orthogonal beam to cover the entire cavity mode volume. Such a beam would inevitably sample many uncoupled molecules; modulation and lock-in schemes would be likely be necessary to correlate off-axis signals with that from a beam aligned through the cavity. Relatedly, the polarizations of on- and off- cavity axis beams must be chosen carefully to address molecules of the same orientational class. In any event, this additional orthogonal spatial degree of freedom provides a wide space to explore in future work.

*Photonic mode engineering.* Engineering the photonic mode structure of gas-phase FP cavities is another untapped degree of freedom. Cavity-altered chemistry in planar microcavities appears to emerge only when a cavity mode with $k = 0$ in-plane momentum is strongly-coupled to a transition of reactant molecules.[53,63,137,138] The significance of this normal-incidence resonance condition is unknown and remains an open area of experimental and theoretical work. It is unclear the extent to which the chemical and photophysical behavior of polaritonic systems arises from the continuous dispersion of modes in a tilt-tuned planar FP microcavity. In contrast, our plano-concave FP cavities support *discrete* Gaussian transverse spatial modes, whose spectrum can be tuned by changing the ratio between the mirror radius of curvature and the cavity length.[20] The ability to manipulate the photonic density of states sets this platform apart from microfluidic cavities and may prove useful to test connections between mode dispersion and cavity-altered chemistry.

*Homogeneous and inhomogeneous broadening.* There are also open questions surrounding the significance of inhomogeneous broadening and molecular disorder to polariton formation, photophysics, and chemistry.[8,13,73,139–142] Cavity transmission spectra and polariton dynamics are sensitive to whether the intracavity molecules chiefly exhibit homogeneous or inhomogeneous broadening. The tunability of gas-phase molecular line shapes is a powerful control knob to enable further investigation into this area: one can explore the interplay between inhomogeneous and homogeneous regimes by tuning the relative contributions of Gaussian Doppler broadening and Lorentzian pressure or transit-time broadening. One could then systematically probe the role of inhomogeneity on polariton dynamics, including the chemical processes discussed above in Sec. IV, and nonlinear spectroscopy (*vide infra*). In addition, future gas-phase work might test Herrera and Owrutsky's prediction that strong coupling is accessible in the pressure-broadened regime for any gas whose absorption cross section exceeds its pressure broadening coefficient.[13]

*Nonlinear spectroscopy.* The gas phase is also a promising arena in which to solidify understanding of the nonlinear spectroscopy of polaritons. Dunkelberger,[52,143] Xiong,[55,144] and Kubarych[145] and coworkers have made major strides in studying the ultrafast optical response of solution-phase vibrational polaritons – though not without some healthy debate.[145,146] In parallel, a growing body of work from Scholes, Zanni, and others examines the nonlinear response of systems under electronic strong coupling.[147–149] Nonlinear spectroscopy of gas-phase polaritons would provide complementary information to these condensed-phase experiments with well-resolved spectral features, non-overlapping excited state absorption signals, and the possibility of directly addressing the "dark state" reservoir off-axis. Altogether, many outstanding questions related to the intrinsic properties and optical response of vibrational and electronic polaritons can be directly probed in the gas phase and in turn inform spectroscopic measures of polariton reactivity.

We must note that there are significant challenges to working with polaritons in the gas phase as well. Absolute Rabi splittings are four to five orders of magnitude smaller for dilute gas samples than those typical in solution-phase microcavities. Centimeter-scale cavities also contain a large number of cavity-coupled molecules (e.g. $N \sim 10^{12}$ for $N/V = 3.5 \times 10^{15}$ cm$^{-3}$) so the effective coupling-per-molecule is orders of magnitude smaller than that of most microcavity implementations. The coupling-per-molecule could be increased by working with smaller mode volumes, more akin to geometries used in few-atom cQED experiments.[30,31] Future work should examine whether moving in this direction is necessary to reveal cavity modification of gas-phase chemistry.

Finally, we note that spectroscopy of molecular gases in optical cavities is by no means a new development: the cavity-enhanced spectroscopy (CES) community has long used cavities to enhance light-matter interactions for more sensitive molecular absorption spectroscopy.[150] A deeper connection to the frameworks developed by the CES community would be of great benefit to researchers in the polariton field. One distinguishing factor of the CES and polaritonic regimes is that in the latter, a photon emitted by a molecule into the cavity mode is more likely to be reabsorbed by the molecular ensemble than to escape the cavity for detection. Experiments that systematically probe chemical reactivity and photophysics at the threshold of strong coupling are of utmost importance to clarify when polaritonic phenomena exceed the grasp of the classical optics language of CES. Gas-phase molecules again make a compelling platform with which to explore this threshold.

In summary, gas-phase molecular strong coupling may provide a powerful means to develop mechanistic understanding of polariton chemistry, harnessing pristine cavity-molecule coupling conditions and state-resolved detection schemes. Drawing inspiration from atomic cavity quantum electrodynamics, cavity-enhanced spectroscopy, and mode-specific chemistry, future work in the gas phase can build toward the dream of optical control of chemistry.

**Supplementary Material**

See the supplementary material for additional details and strong coupling simulations. In Sec. SI, we define absorption line shape broadening and tabulate specific broadening coefficients for all molecular systems considered in this work (Table SI). In Sec. SII, Figures S1 and S2 depict absorption cross sections and strong coupling prospects for RVSC of $CH_3OH$ and ESC of $O_3$. In Sec. SIII, we explicitly consider the temperature dependence of the HONO *trans-cis* isomer ratio.

## Acknowledgements

Marissa L. Weichman acknowledges support from NSF CAREER award CHE-2238865 and startup funds provided by Princeton University. Jane C. Nelson acknowledges support from a National Defense Science and Engineering Graduate fellowship. The authors acknowledge helpful conversations with David Chandler and Jeff Owrutsky.

## Author Declarations

*Conflict of Interest*

The authors have no conflicts to disclose.

*Author Contributions*

Jane C. Nelson: Conceptualization; Data curation; Formal analysis; Investigation; Methodology; Visualization; Writing – original draft; Writing – review & editing

Marissa L. Weichman: Conceptualization; Funding acquisition; Methodology; Project administration; Resources; Supervision; Writing – review & editing

*Data availability*

The data that support the findings of this study are available from the corresponding author upon reasonable request.

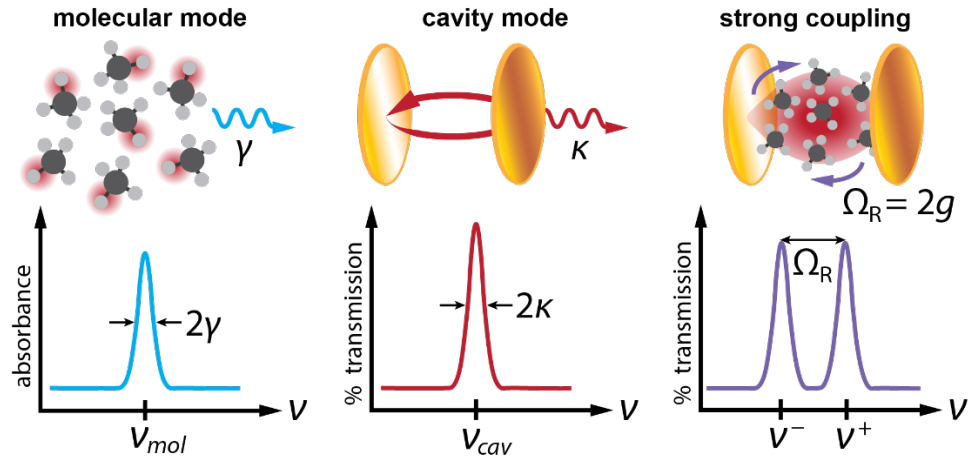

**FIG. 1.** Schematic of strong light-matter coupling when a molecular excitation frequency ($\nu_{mol}$) is resonant with an optical cavity mode frequency ($\nu_{cav}$). When the rate of photon exchange between molecules and cavity exceeds both the molecular relaxation rate, $\gamma$, and the cavity photon loss rate, $\kappa$, the light-matter system enters the strong coupling regime. In this regime, cavity transmission maxima appear at new frequencies, $\nu^{\pm}$, separated by the Rabi frequency, $\Omega_R$.

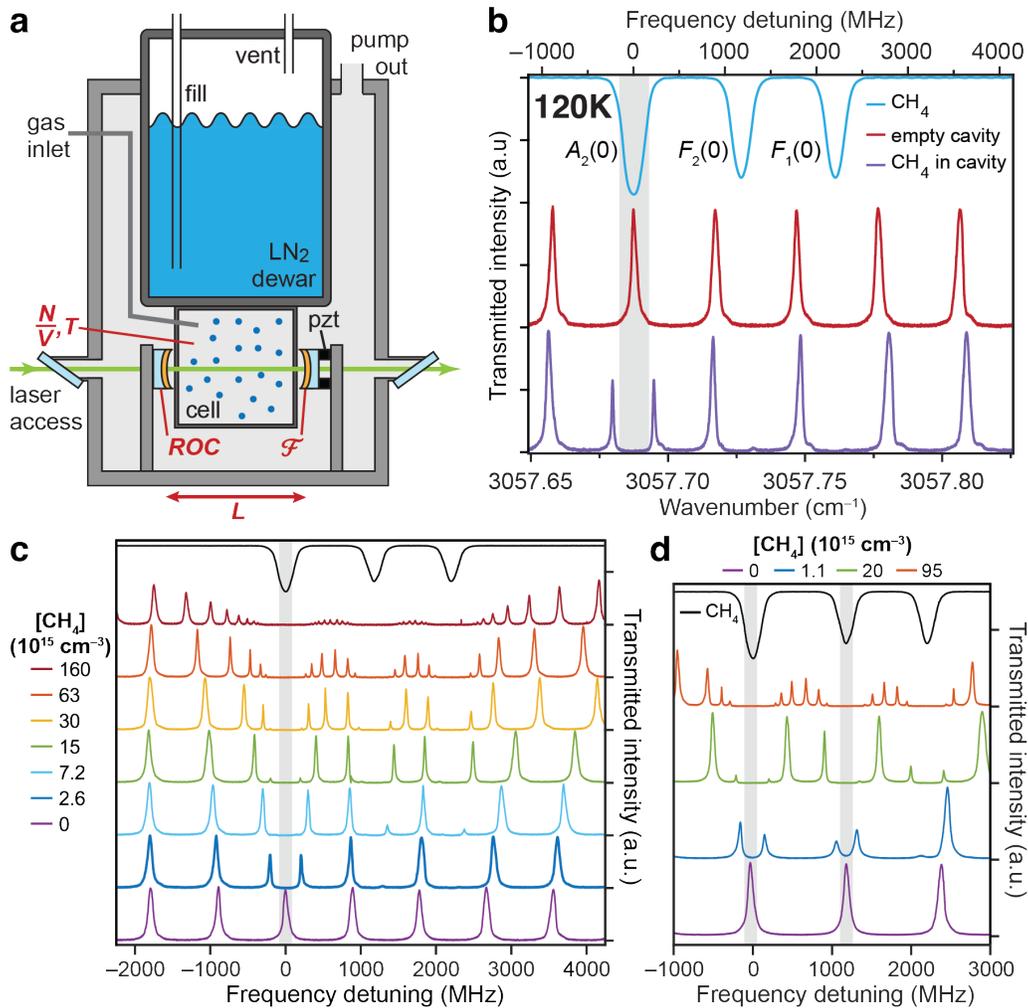

**FIG. 2. (a)** Intracavity cryogenic buffer gas cell (CBGC) used to achieve strong coupling with gas-phase molecules of varying number density ($N/V$) and temperature ($T$). The Fabry-Pérot (FP) cavity geometry is determined by its length ($L$), finesse ($\mathcal{F}$), and the mirror radius of curvature (ROC). **(b)** Transmission spectra of methane ($CH_4$, blue), an empty FP cavity (red), and methane in the same cavity under strong coupling conditions for the target $\nu_3$, $J = 3\rightarrow 4$, $A_2(0)$ transition (purple). Methane spectra are acquired at 120 K with $N/V = 1.5\times 10^{15}$ cm$^{-3}$ for the extracavity sample and $N/V = 3.5\times 10^{15}$ cm$^{-3}$ for the strongly-coupled sample. The near-confocal cavity used here has $L = 8.36$ cm, $\mathcal{F} = 24$, and ROC $= -8.36$ cm. The empty cavity sustains a free spectral range of 895 MHz (0.0299 cm$^{-1}$) and a linewidth of 65 MHz fwhm (0.0022 cm$^{-1}$). A Rabi splitting of 454 MHz (0.0151 cm$^{-1}$) is achieved under these conditions. **(c)** Transmission spectra of the same cavity from panel (b) containing increasing intracavity number densities of methane at 120 K. **(d)** Transmission spectra of a near-confocal $L = 6.27$ cm, $\mathcal{F} = 25$ cavity containing increasing intracavity methane number densities. Adjacent cavity modes are near-resonant with the $\nu_3$, $J = 3\rightarrow 4$ $A_2(0)$ and $F_2(0)$ transitions so that strong coupling is achieved with two transitions simultaneously. Panels (a) and (b) are adapted with permission from Wright *et al.*, J. Am. Chem. Soc. 145, 5982 (2023). Copyright 2023 American Chemical Society.[10] Panels (c), and (d) are adapted with permission from Wright *et al.*, J. Chem. Phys. 159, 164202 (2023). Copyright 2023, AIP Publishing.[11]

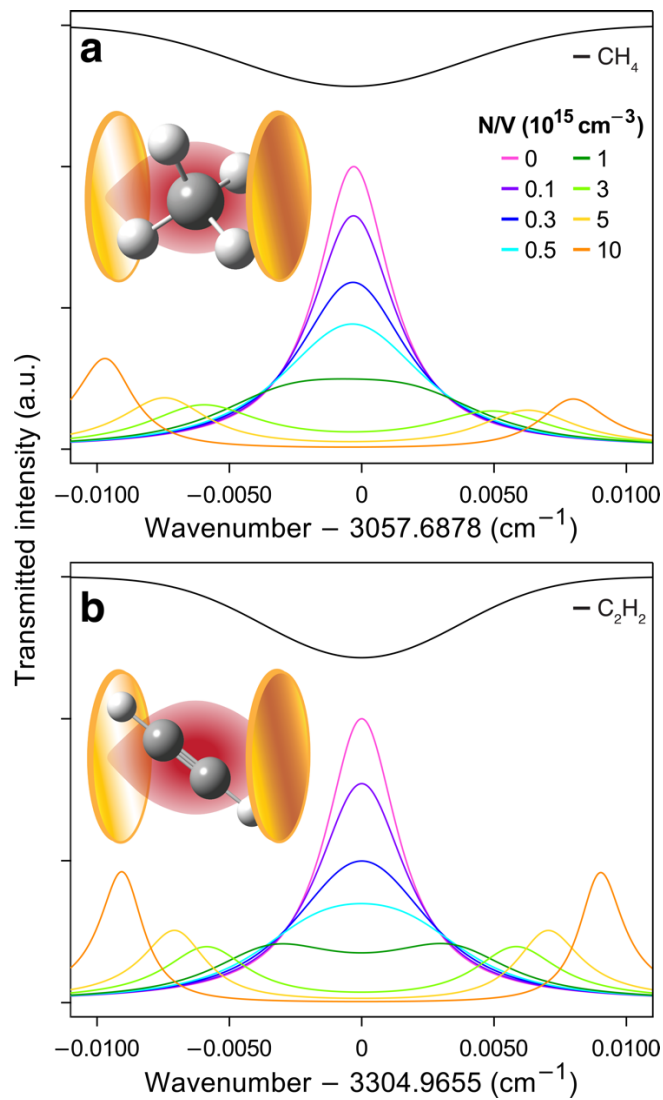

**FIG. 3.** Simulated classical cavity transmission spectra through an $L = 5$ cm, $\mathcal{F} = 30$ cavity under strong coupling of individual rovibrational transitions of **(a)** methane (CH$_4$, $\nu_3$, $J = 3\rightarrow4$) and **(b)** acetylene (C$_2$H$_2$, $\nu_3$, $J = 9\rightarrow10$) at 295 K and varying number densities (colored traces). Simulated transmission spectra through 5 cm pathlengths of **(a)** CH$_4$ and **(b)** C$_2$H$_2$ at $N/V = 1\times10^{15}$ cm$^{-3}$ and 295 K are shown in black and offset for clarity. Molecular absorption cross sections are simulated using HITRAN line list parameters in PGOPHER with appropriate Doppler and pressure broadening.

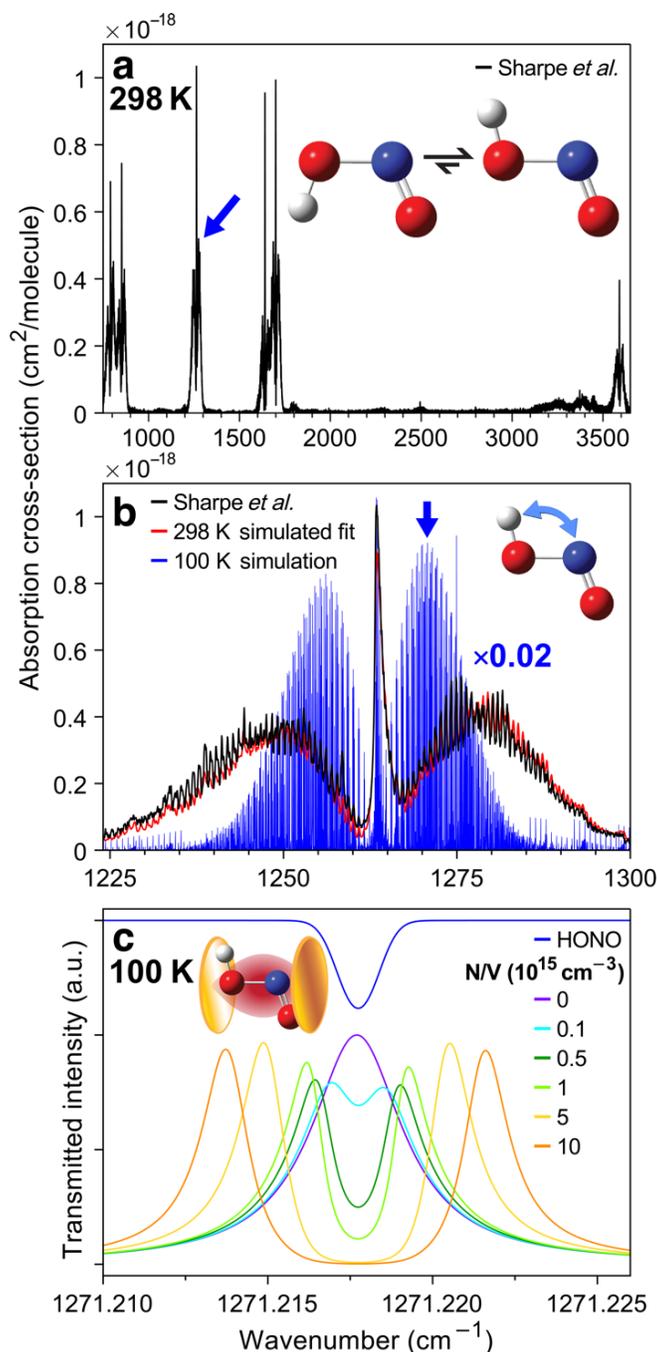

FIG. 4. (a) Broadband infrared absorption cross section of nitrous acid (HONO) at 298 K from Sharpe *et al.*[106] obtained via HITRAN. (b) Absorption cross section of the $\nu_3$ HON bending mode of *trans*-HONO. The 298 K experimental absorption cross section (black) is fit with a PGOPHER simulation using spectroscopic constants from Gaussian calculations and literature (red). The PGOPHER model is then used to calculate the low-temperature, low-pressure absorption cross section at 100 K and 0.0104 torr ($N/V \sim 1 \times 10^{15}$ cm$^{-3}$), which is scaled by 0.02 for plotting (blue). (c) Classical cavity transmission simulation for coupling a $\nu_3$, $J = 9 \rightarrow 10$ rovibrational transition of *trans*-HONO in a $L = 5$ cm, $\mathcal{F} = 30$ cavity for varying intracavity number densities at 100 K (colored traces). A simulated trace of transmission through a 5 cm pathlength of an $N/V = 1 \times 10^{15}$ cm$^{-3}$, 100 K sample of *trans*-HONO is offset for clarity (blue).

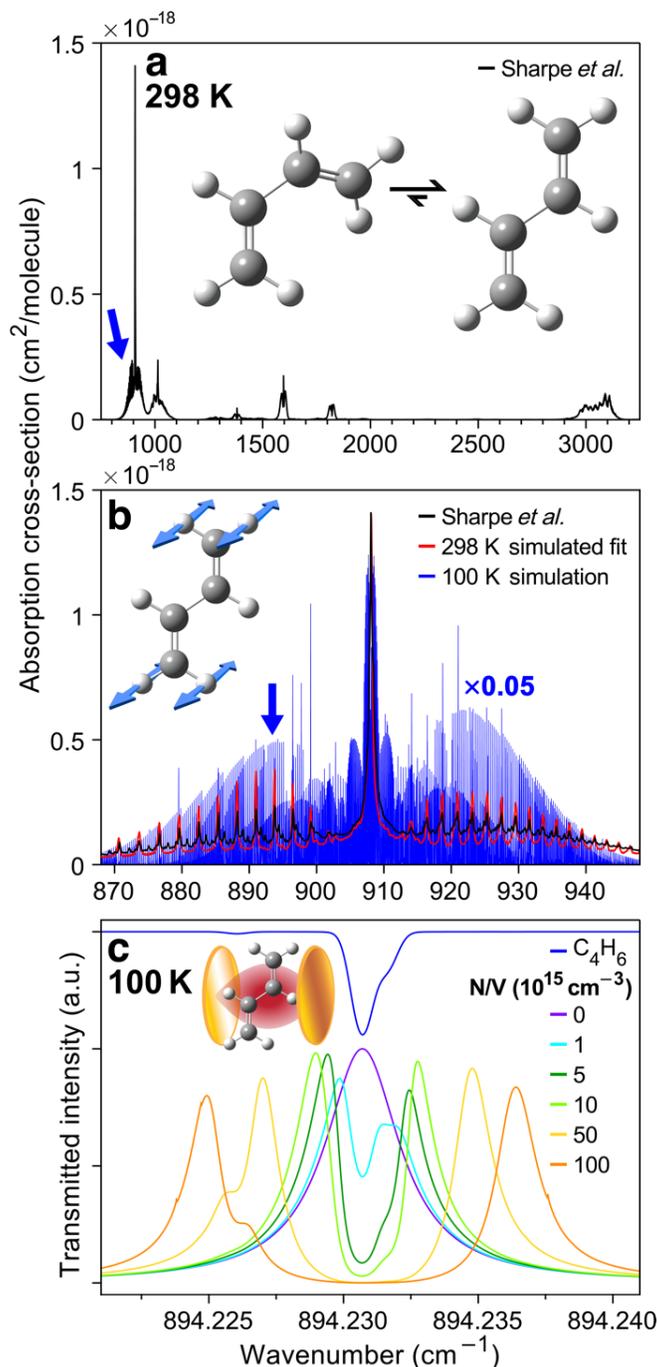

**FIG. 5. (a)** Broadband infrared absorption cross section of butadiene ($C_4H_6$) at 298 K from Sharpe *et al.*[106] obtained via HITRAN. **(b)** Absorption cross section of the $\nu_{11}$ $CH_2$ wagging mode of *trans*-$C_4H_6$. The 298 K experimental absorption cross section (black) is fit with a PGOPHER simulation using spectroscopic constants from Gaussian calculations and literature (red). The PGOPHER model is then used to calculate the low-temperature, low-pressure absorption cross section at 100 K and 0.0104 torr ($N/V \sim 1\times10^{15}$ cm$^{-3}$), which is scaled by 0.05 for plotting (blue). **(c)** Classical cavity transmission simulation for coupling overlapping $\nu_{11}$, $J = 8\rightarrow7$ and $J = 30\rightarrow29$ transitions of *trans*-$C_4H_6$ in a $L = 5$ cm, $\mathcal{F} = 30$ cavity for varying intracavity number densities at 100 K (colored traces). A simulated trace of transmission through a 5 cm pathlength of an $N/V = 5\times10^{15}$ cm$^{-3}$, 100 K sample of *trans*-$C_4H_6$ is offset for clarity (blue).

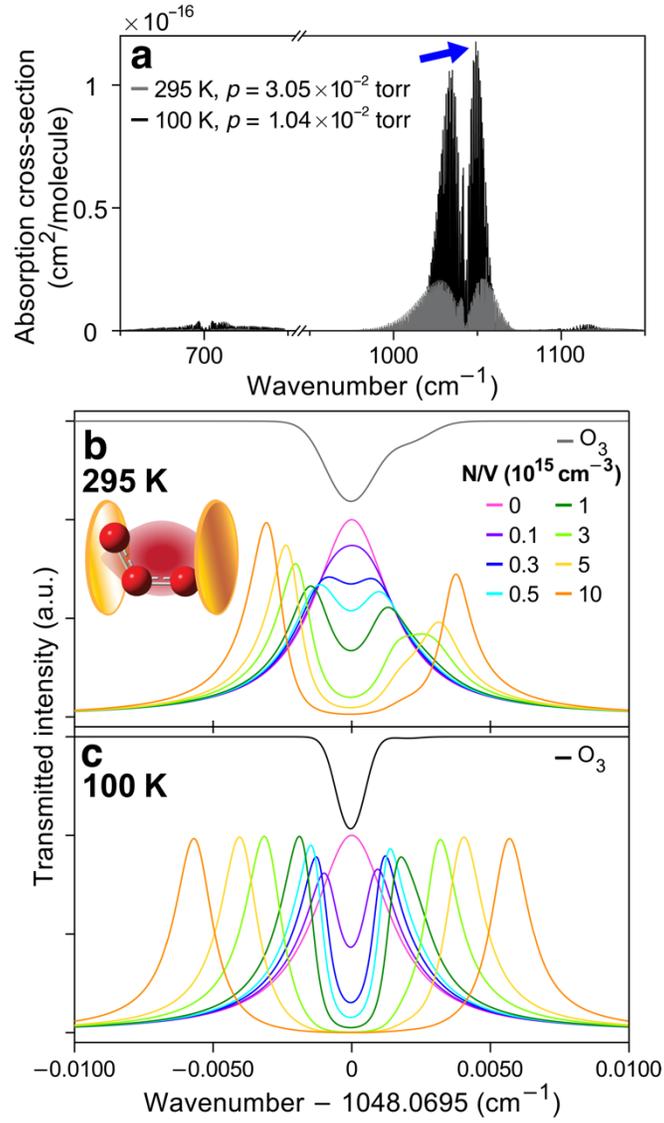

**FIG. 6. (a)** Broadband infrared absorption cross section of ozone ($O_3$) at 295 K (gray) and 100 K (black) with pressures corresponding to molecular densities of $N/V = 1 \times 10^{15}$ cm$^{-3}$ simulated using HITRAN line list parameters in PGOPHER. **(b, c)** Classical cavity transmission simulations for coupling a $\nu_3$, $J = 7 \rightarrow 8$ rovibrational transition of $O_3$ in a $L = 5$ cm, $\mathcal{F} = 30$ cavity for varying intracavity number densities at **(b)** 295 K and **(c)** 100 K (colored traces). Simulated traces of transmission through a 5 cm pathlength of **(b)** 295 K, $N/V = 3 \times 10^{15}$ cm$^{-3}$ and **(c)** 100 K, $N/V = 5 \times 10^{14}$ cm$^{-3}$ samples of $O_3$ are shown in gray and black, respectively, and offset for clarity.

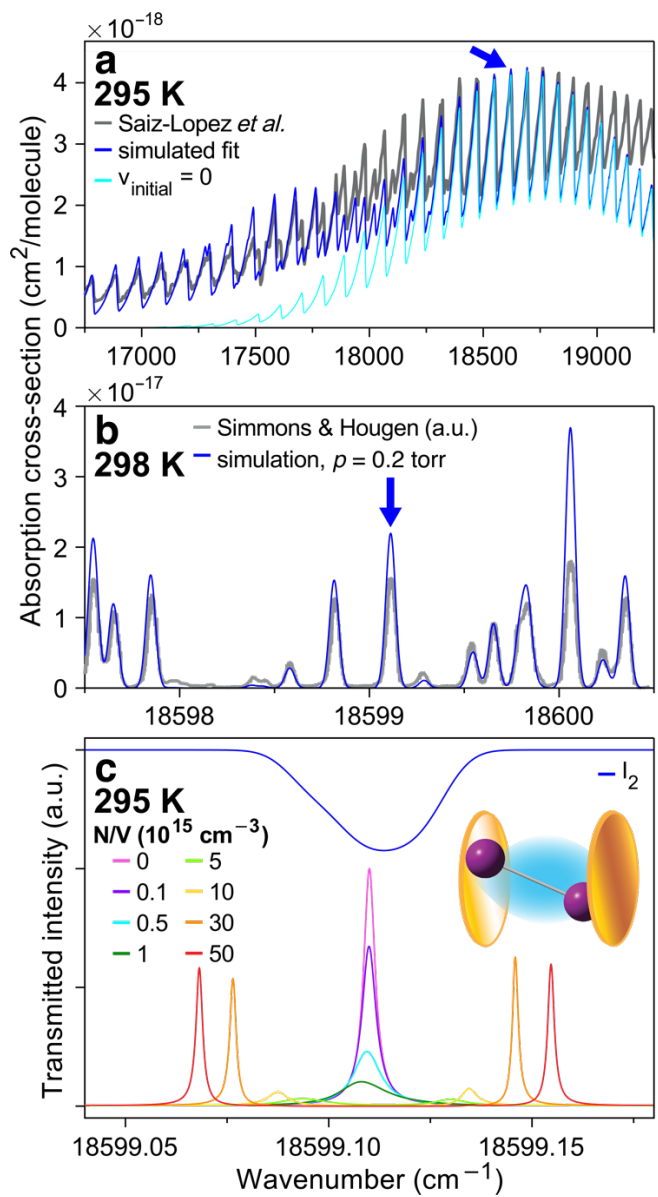

**FIG. 7. (a)** Absorption cross section of the B−X electronic band of iodine ($I_2$) at 295 K. Experimental air-broadened data from Saiz-Lopez et al.[134] (gray) is fit with a PGOPHER model built from literature constants (blue). Contributions from transitions originating in the $v_{initial} = 0$ level of the electronic ground state are shown in cyan. **(b)** The $I_2$ model (blue) reproduces rovibronic line positions of experimental absorbance data digitized from Simmons and Hougen (gray),[128] assuming a Gaussian linewidth of 0.055 cm$^{-1}$. The reference data from Simmons and Hougen is reported with arbitrary intensity units and strong transitions may be saturated. **(c)** Classical cavity transmission simulation for coupling the B−X, $v_1 = 0 \rightarrow 29$, $J = 45 \rightarrow 46$ rovibronic transition of $I_2$ in a $L = 5$ cm, $\mathcal{F} = 30$ cavity for varying intracavity number densities at 295 K (colored traces). A simulated trace of transmission through a 5 cm pathlength of an $N/V = 1 \times 10^{15}$ cm$^{-3}$, 295 K sample of $I_2$ is offset for clarity (blue). The asymmetric line shape of the rovibronic transition arises from unresolved hyperfine structure.

**TABLE I.** Conditions for strong coupling simulations of molecular gases.

| Molecule | Transition | $T$ (K) | Molecular linewidth (fwhm, cm$^{-1}$) [a] | $N/V$ (cm$^{-3}$) [b] | $\Omega_R$ (cm$^{-1}$) [c] |
|---|---|---|---|---|---|
| CH$_4$ | $\nu_3$, $J = 3 \to 4$ 3057.6878 cm$^{-1}$ | 295 | 0.00940 | $3\times10^{15}$ | 0.01087 |
| C$_2$H$_2$ | $\nu_3$, $J = 9 \to 10$ 3304.9655 cm$^{-1}$ | 295 | 0.00799 | $3\times10^{15}$ | 0.01170 |
| CH$_3$OH | $\nu_8$, $J = 9 \to 10$ 1049.3578 cm$^{-1}$ | 295 | 0.00244 | $9\times10^{15}$ | 0.00358 |
|  |  | 100 | 0.00136 | $5\times10^{15}$ | 0.00340 |
| trans-HONO | $\nu_3$, $J = 9 \to 10$ 1271.218 cm$^{-1}$ | 100 | 0.00135 | $3\times10^{15}$ | 0.00449 |
| trans-C$_4$H$_6$ | $\nu_{11}$, $J = 8 \to 7$ 894.2307 cm$^{-1}$ $\nu_{11}$, $J = 30 \to 29$ 894.2316 cm$^{-1}$ | 100 | 0.00101[d] | $1\times10^{16}$ | 0.00378 |
| O$_3$ | $\nu_3$, $J = 7 \to 8$ 1048.0695 cm$^{-1}$ | 295 | 0.00199 [d] | $3\times10^{15}$ | 0.00456 |
|  |  | 100 | 0.00108 | $9\times10^{14}$ | 0.00351 |
| I$_2$ | B–X, $v_1 = 0 \to 29$, $J = 45 \to 46$ 18599.1 cm$^{-1}$, 538 nm | 295 | 0.02844 | $5\times10^{15}$ | 0.03647 |
| O$_3$ [e] | B–X 39300 cm$^{-1}$, 254 nm | 293 | 6552 | $3\times10^{21}$ | 7027 |

[a] The molecular linewidth reflects the Voigt profile of the simulated absorption coefficient and therefore represents a convolution of both Gaussian Doppler broadening and Lorentzian pressure broadening.
[b] The number density given is the smallest value simulated that reaches the strong coupling regime, as defined by the Rabi splitting exceeding both molecular transition and cavity mode fwhm linewidths.
[c] Classical cavity transmission simulations were performed with a non-confocal $L = 5$ cm, $\mathcal{F} = 30$ cavity with a $\Delta\nu = 0.00336$ cm$^{-1}$ fwhm linewidth unless otherwise noted.
[d] These systems feature asymmetric molecular line shapes with distinct shoulders not captured by the molecular linewidth. In these cases, the strong coupling regime is determined when the Rabi splitting exceeds the cavity linewidth.
[e] Classical cavity transmission simulations for the O$_3$ B–X band were performed with an $L = 254$ nm, $\mathcal{F} = 6$ cavity with a $\Delta\nu = 3267$ cm$^{-1}$ fwhm cavity linewidth. The absorption cross section data used in transmissions simulations are from Serdyuchenko et al.[136] rather than PGOPHER simulations.

# Supplementary Material:
## *More than just smoke and mirrors*:
## Gas-phase polaritons for optical control of chemistry

*Jane C. Nelson, Marissa L. Weichman*[*]

Department of Chemistry, Princeton University, Princeton, New Jersey, 08544, USA
[*] weichman@princeton.edu

**SI. Broadening conditions for simulated molecular absorption cross sections**

Molecular absorption cross sections are sensitive to environmental conditions which cause homogeneous and inhomogeneous broadening. We simulate Voigt absorption line shapes arising from the convolution of Gaussian (Doppler or instrumental) broadening and Lorentzian (pressure) broadening under conditions relevant to our experiments. We neglect transit time broadening, which is comparably small under our experimental conditions.[1]

To treat Doppler broadening, we calculate the Gaussian full-width at half-maximum (fwhm) linewidth for each molecular transition according to

$$\Delta \nu_G = \sqrt{8 \ln(2)} \cdot \nu_0 \cdot \sqrt{\frac{k_B T}{mc^2}} \qquad (S1)$$

where $\nu_0$ is the central transition frequency, $T$ is the translational temperature, $m$ is the molecular mass, $c$ is the speed of light, and $k_B$ is the Boltzmann constant.

We calculate the Lorentzian fwhm linewidth from pressure broadening according to

$$\Delta \nu_L = 2 \gamma p \qquad (S2)$$

where $p = N k_B T / V$ is the pressure determined using the ideal gas law and $\gamma$ is the relevant self-broadened or air-broadened half-width at half-maximum coefficient for a given molecular transition which is obtained either from the HITRAN database or the literature (see Table SI). Further discussion of pressure broadening in mixed gas samples can be found in the HITRAN documentation.[2] We neglect the temperature dependence of pressure broadening coefficients as the pressure broadening coefficients and their temperature dependence have large uncertainties for many of the species considered here. Moreover, Doppler broadening dominates our simulated molecular linewidths by at least an order of magnitude even at low temperatures. Therefore, we do not expect the temperature dependence of pressure broadening to change our predictions about strong coupling prospects and necessary number densities.

In Table SI, we tabulate representative broadening conditions used in this work to simulate absorption cross sections in PGOPHER.

**TABLE SI.** Experimental conditions and broadening parameters used to fit and simulate absorption cross sections of molecular gases.

| Molecule | $\nu_0$ (cm$^{-1}$) | $T$ (K) | $N/V$ (cm$^{-3}$) [a] | $p$ (torr, atm) | $\gamma_{self}$ (cm$^{-1}$/atm) | $\Delta\nu_G$ (fwhm, cm$^{-1}$) | $\Delta\nu_L$ (fwhm, cm$^{-1}$) |
|---|---|---|---|---|---|---|---|
| CH$_4$ | 3058 | 295 | 3×10$^{15}$ | 0.0916, 1.21×10$^{-4}$ | 0.079 [b] | 9.39×10$^{-3}$ | 1.90×10$^{-5}$ |
| C$_2$H$_2$ | 3305 | 295 | 3×10$^{15}$ | 0.0916, 1.21×10$^{-4}$ | 0.154 [b] | 7.97×10$^{-3}$ | 3.71×10$^{-5}$ |
| CH$_3$OH | 1049 | 295 | 9×10$^{15}$ | 0.275, 3.62×10$^{-4}$ | 0.4 [b] | 2.28×10$^{-3}$ | 2.89×10$^{-4}$ |
| CH$_3$OH | 1049 | 100 | 5×10$^{15}$ | 0.0518, 6.81×10$^{-5}$ | 0.4 [b] | 1.33×10$^{-3}$ | 5.45×10$^{-5}$ |
| trans-HONO | 1271 | 298 [c] | | | | 0.12 | 0.28 |
| trans-HONO | 1271 | 100 | 3×10$^{15}$ | 0.0311, 4.09×10$^{-5}$ | 0.4 [d] | 1.33×10$^{-3}$ | 3.27×10$^{-5}$ |
| trans-C$_4$H$_6$ | 894 | 298 [c] | | | | 0.12 | 0.22 |
| trans-C$_4$H$_6$ | 894 | 100 | 1×10$^{16}$ | 0.104, 1.36×10$^{-4}$ | 0.1 [e] | 8.71×10$^{-4}$ | 2.74×10$^{-5}$ |
| O$_3$ | 1048 | 295 | 3×10$^{15}$ | 0.0916, 1.21×10$^{-4}$ | 0.1 [b] | 1.86×10$^{-3}$ | 2.41×10$^{-5}$ |
| O$_3$ | 1048 | 100 | 9×10$^{14}$ | 0.00932, 1.23×10$^{-5}$ | 0.1 [b] | 1.08×10$^{-3}$ | 2.45×10$^{-6}$ |
| I$_2$ | 18599 (538 nm) | 295 [f] | | | | 5 | 0.4 |
| I$_2$ | 18599 (538 nm) | 298 [g] | | 0.2, 2.6×10$^{-4}$ | 0.2 [h] | 5.5×10$^{-2}$ | 1×10$^{-4}$ |
| I$_2$ | 18599 (538 nm) | 295 | 5×10$^{15}$ | 0.153, 2.01×10$^{-4}$ | 0.2 [h] | 1.44×10$^{-2}$ | 8.04×10$^{-5}$ |

[a] The given $N/V$ corresponds to the lowest molecular number density for which the onset of the strong coupling regime is observed. The onset of strong coupling is defined when the Rabi splitting exceeds both molecular transition and cavity mode fwhm linewidths.
[b] Self-broadening coefficient obtained from HITRAN.[2]
[c] Conditions used to fit room temperature reference data from Sharpe et al.[3] who report an instrumental resolution of 0.112 cm$^{-1}$ and prepare samples in 760 torr of N$_2$. The fitted Gaussian fwhm is consistent with reported instrumental resolution and the fitted Lorentzian fwhm falls in a range consistent with pressure broadening of the target species.
[d] Self-broadening coefficient obtained from Armante et al.[4]
[e] Self-broadening coefficient estimated from literature and comparison to similar species.[2,5–9]
[f] Conditions used to fit room temperature reference data from Saiz-Lopez et al.[10] who report an instrumental resolution of 4 cm$^{-1}$ and prepare I$_2$ samples in 760 torr of air.
[g] Conditions used to fit room temperature reference data from Simmons and Hougen[11] who report an experimental linewidth of 0.055 cm$^{-1}$ and use an I$_2$ sample near its vapor pressure (0.2 torr).[10]
[h] Self-broadening coefficient estimated from literature.[12–17]

## SII. Additional simulations of molecular cavity-coupling

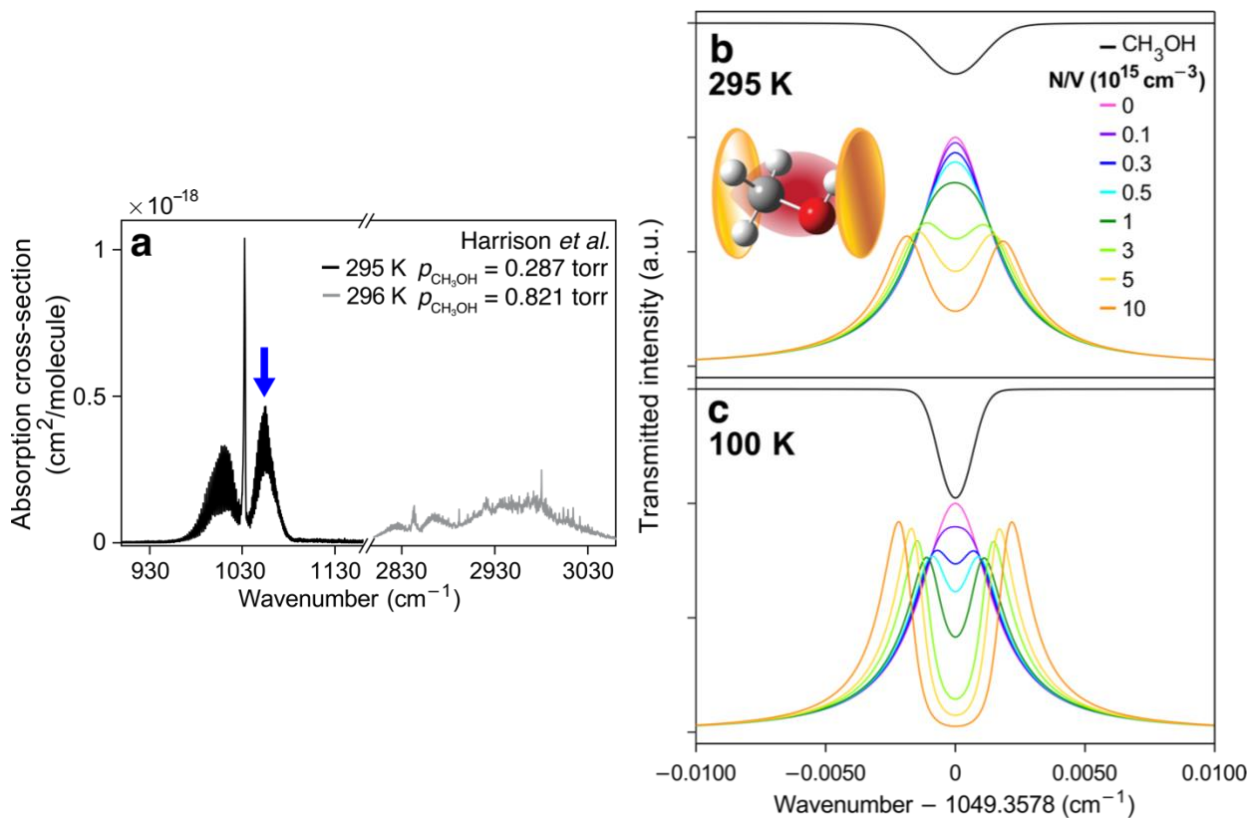

**FIG. S1. (a)** Broadband infrared absorption cross section of methanol ($CH_3OH$) near room temperature from Harrison *et al.*[18] obtained via HITRAN. Reference data correspond to methanol partial pressures of 0.287 torr in the spectral region near 10 μm (black) and 0.821 torr near 3.4 μm (gray), pressure-broadened in ~761 torr of air. **(b, c)** Classical cavity transmission simulations for coupling a $\nu_8$, $J = 9 \rightarrow 10$ rovibrational transition of $CH_3OH$ in a $L = 5$ cm, $\mathcal{F} = 30$ cavity at **(b)** 295 and **(c)** 100 K (colored traces). A HITRAN line list is available for the 10 μm region of $CH_3OH$ and is used to simulate the absorption cross section at relevant temperature and pressure conditions for varying number densities. Simulated traces of the transmission through a 5 cm pathlength of **(b)** 295 K, $N/V = 1 \times 10^{16}$ cm$^{-3}$ and **(c)** 100 K, $N/V = 5 \times 10^{15}$ cm$^{-3}$ samples of $CH_3OH$ are offset for clarity (black).

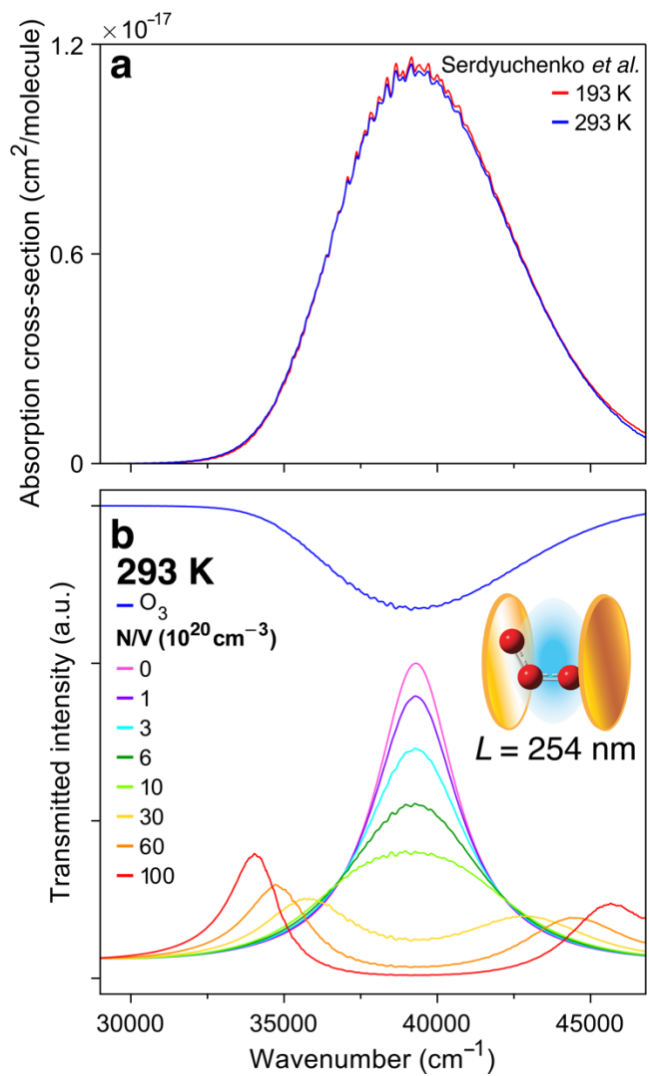

**FIG. S2. (a)** Absorption cross section of the ozone ($O_3$) B−X Hartley band from Serdyuchenko *et al.*[19] at 193 K (red) and 293 K (blue) near 250 nm. **(b)** Classical cavity transmission simulations for coupling the entire $O_3$ B−X band in a $L$ = 254 nm, $\mathcal{F}$ = 6 finesse cavity. Simulations are performed using the 293 K cross section data for varying molecular number densities (colored traces). A simulated trace of the transmission (blue) through a 254 nm pathlength of a $N/V = 6\times10^{20}$ cm$^{-3}$ sample of $O_3$ is offset for clarity (blue).

S4

**SIII. The temperature-dependent *trans-cis* isomeric ratio of HONO**

The temperature-dependent *trans*:*cis* ratio of gas-phase HONO is not well-characterized because the energy gap between isomers has large uncertainty.[4] The *trans* isomer is known to be the thermodynamic ground state; the *trans*:*cis* ratio has been reported as 1.7:1 and 2.3:1 for room temperature gas;[4,20,21] 3.3:1 for 277 K gas;[22] and as high as 6:1 in a 10 K inert gas matrix.[23] The increasing population of *trans*-HONO at low temperatures will increase its line strengths and impact the accuracy of the strong coupling simulations discussed in Sec. IV B and Fig. 4 of the main text. In particular, strong coupling may be accessible at slightly lower *trans*-HONO number densities than predicted in the main text.

Here, we provide an estimate of the impact of these thermodynamic considerations. We assume a thermalized Boltzman distribution, with

$$r_{tc}(T) = \frac{N_t(T)}{N_c(T)} = \exp(-\Delta E_{tc}/k_B T) \tag{S3}$$

where $r_{tc}(T)$ is the temperature-dependent *trans*:*cis* ratio, $N_{t,c}(T)$ is the absolute number of molecules in either the *trans* or *cis* configuration, and $\Delta E_{tc}$ is the *trans-cis* energy difference. At the high end of room-temperature isomeric ratios, Barney *et al.*[20] report $r_{tc}(296 \text{ K}) = 2.3$ corresponding to $\Delta E_{tc} = -171$ cm$^{-1}$. Using this value for the energy gap, we find $r_{tc}(100 \text{ K}) = 12$. Cooling from room temperature to 100 K will increase the *trans*:*cis* ratio by a factor of 12/2.3 = 5.2. We therefore estimate that the *trans* isomer will make up 70% of the population at room temperature and 92% at 100 K, so *trans*-HONO transitions will increase in intensity by *at most* 30% upon cooling. The temperature-dependent isomeric equilibrium therefore does not overcome the two orders of magnitude gap between experimentally accessible HONO number densities and those needed to reach the strong coupling regime.

During the preparation of this manuscript, a new high-resolution IR spectroscopy study of HONO was published[24] that supports a smaller room-temperature *trans-cis* energy difference of $\Delta E_{tc} = -96$ cm$^{-1}$. This data point further constrains the expected intensity increase of *trans* features upon cooling.